# Cascading LO-phonon-mediated resonant energy transfer from excitons to $Eu^{3+}$ for ultrahigh color purity red luminescence in europium-doped GaN

**Authors:** Shizhao Fan,[1,2,*] Yue Xu,[3] Yuhao Yin,[2] Shitao Dong,[3] Shaoqiang Chen,[4] Xueyuan Chen,[5] Zhan Su,[4] Datao Tu,[5] Qian Sun,[1,2,†] Hui Yang,[1,2,]

[1] Key laboratory of Semiconductor Display Materials and Chips, Suzhou Institute of Nano-Tech and Nano-Bionics (SINANO), Chinese Academy of Sciences (CAS), Suzhou 215123, P. R. China

[2] Division of Nano-Devices Research, SINANO, CAS, Suzhou 215123, P. R. China

[3] School of Nano Science and Technology, University of Science and Technology of China, Suzhou 215123, P. R. China

[4] State Key Laboratory of Precision Spectroscopy, Department of Electronic Engineering, East China Normal University, Shanghai 200241, P. R. China

[5] State Key Laboratory of Structural Chemistry and Fujian Key Laboratory of Nanomaterials, Fujian Institute of Research on the Structure of Matter, CAS, Fuzhou, China

*Contact author: szfan2020@sinano.ac.cn
[†]Contact author: qsun2011@sinano.ac.cn

**ABSTRACT:** High-efficiency, wavelength-stable red micro-LEDs with ultrahigh color purity remain a critical bottleneck for monolithic III-nitride full-color displays. Europium-doped GaN (GaN:Eu) offers a compelling solution, yet the microscopic energy transfer mechanism from the GaN host to $Eu^{3+}$ ions has remained unresolved for three decades, impeding rational performance optimization. Here we report polaron-mediated resonant energy transfer in high-crystalline-quality GaN:Eu epilayers grown by plasma-assisted molecular beam epitaxy. $Eu^{3+}$ doping stabilizes a distinct bound exciton ($BX_3$) state at 3.478 eV, which couples strongly with longitudinal optical (LO) phonons via the Fröhlich interaction to form cascading polarons with up to seven phonon replicas at a uniform energy spacing of 92.0 ±0.7 meV. The $BX_3$ polaron coupled with 5×LO phonons matches precisely the $^7F_0 \rightarrow {}^5D_3$ intra-4*f* transition energy of $Eu^{3+}$ at 3.019 eV. Temperature-dependent photoluminescence reveals non-equilibrium phonon emission, evidencing directional resonant energy transfer. After rapid nonradiative relaxation from the $^5D_3$ excited state to the $^5D_0$ metastable state, the hypersensitive $^5D_0 \rightarrow {}^7F_2$ transition yields sharp 622 nm red emission. Time-resolved spectroscopy confirms a single dominant $Eu^{3+}$ luminescent site with reduced local symmetry, attributed to off-axis $Eu_{Ga}$-$V_N$ complexes. This work identifies a key energy transfer channel and provides a clear framework for designing high-efficiency GaN:Eu red emitter.

## INTRODUCTION

Augmented and virtual reality near-eye displays impose stringent requirements on the resolution, color gamut, and luminous efficiency of micro-LEDs. While substantial progress has been made in blue and green InGaN quantum well micro-LEDs[1-8], the realization of high-resolution monolithic full-color displays is primarily hindered by the low brightness and poor color purity of red micro-LEDs. Commercial AlGaInP-based red LEDs suffer severe efficiency degradation at microscales[9], with sidewall surface recombination velocity one order of magnitude higher than that of III-nitride counterparts[10]. Low thermal conductivity, weak electron confinement, and thermally induced leakage at high injection densities further compromise device efficiency and stability[11], while the mismatched mechanical, chemical, and thermal properties create reliability challenges for monolithic integration with GaN-based pixels. Blue-to-red color conversion via colloidal quantum dots faces comparable limitations including thermal quenching, poor stability, and residual blue light leakage[12, 13]. These hurdles drive strong demand for GaN-based red micro-LEDs to enable monolithic III-nitride RGB integration.

Red emission from $In_xGa_{1-x}N$/GaN quantum wells requires an indium content $x$~0.4, which demands low growth temperatures, introduces substantial compressive strain, and degrades material quality due to the wide InN—GaN miscibility gap[14-16]. Consequently, state-of-the-art InGaN red LEDs exhibit electroluminescence (EL) spectral full-widths-half-maximum (FWHMs) exceeding 50 nm, plus over 30 nm wavelength blueshift as the injection current density increases from <1 A/cm$^2$ to >10 A/cm$^2$, driven by alloy inhomogeneity and quantum-confined Stark effect. To circumvent these issues, intensive research has focused on incorporating trivalent europium ($Eu^{3+}$) ions into the GaN lattice, which yields sharp, intense and temperature-insensitive red emission at 622 nm, originating from the intra-4$f$ $^5D_0 \rightarrow {}^7F_2$ transition[17-19]. Theoretical and experimental studies confirm that isovalent lanthanide ion substitution at Ga sites ($RE_{Ga}$) is thermodynamically stable, and that $RE_{Ga}$-$V_N$ ($V_N$, nitrogen vacancy) pairs are the most promising luminescent center candidates[20, 21].

Since the first demonstration of EL from GaN:Eu three decades ago[22, 23], strategies including ion implantation, epitaxial co-doping, and active-region engineering have been developed to improve luminescence intensity. Eu-O co-doping[18, 24] has yielded external quantum efficiencies up to 9.2%[19], and growth on semipolar GaN has tripled photoluminescence (PL) intensity[25]. Eight distinct photoactive $Eu^{3+}$ sites have been identified by combined excitation-emission spectroscopy (CEES)[26-29]. Despite this progress, the fundamental mechanism of energy transfer from the GaN host to photoactive $Eu^{3+}$ ions remains elusive, fundamentally limiting further device optimization. Early work employed a defect-related Auger excitation model[19, 30], while subsequent studies proposed that point defects trap electron-hole pairs adjacent to $Eu^{3+}$ [26, 31], and that free carriers from thermally dissociated excitons drive luminescence[32]. In contrast, a pronounced temperature-insensitive excitation peak at the bound-exciton energy confirms direct exciton-to-$^5D_{J'}$ energy transfer[33], though the specific transition pathway and excess energy dissipation mechanism remains unknown[34]. It has long been speculated that energy transfer proceeds via resonant Auger scattering, with excess energy compensated by phonon emission[35], yet no direct spectroscopic evidence of intermediate polaron states has been detected to date.

In this work, we elucidate this long-standing puzzle through observation of cascading bound-exciton–LO-phonon polarons in high-crystalline-quality GaN:Eu epilayers grown by plasma-assisted molecular beam epitaxy (PA-MBE). We identify a distinct bound exciton stabilized by $Eu^{3+}$ doping, which forms a series of polaron states with up to seven well-resolved phonon replicas via the Fröhlich interaction. The $5^{th}$-order polaron state resonates exactly with the $Eu^{3+}$ $^7F_0 \rightarrow {^5D_3}$ transition, enabling efficient resonant energy transfer. Temperature-dependent measurements reveal non-equilibrium phonon dynamics characteristic of directional energy flow, rather than thermal equilibrium phonon sidebands. Combining time-resolved and high-resolution cryogenic spectroscopy further establishes that a single dominant off-axis $Eu_{Ga}$-$V_N$ luminescent site with reduced symmetry governs the red emission. These results elucidate a key energy transfer pathway in GaN:Eu and provide a physical basis for engineering high-efficiency red micro-LEDs.

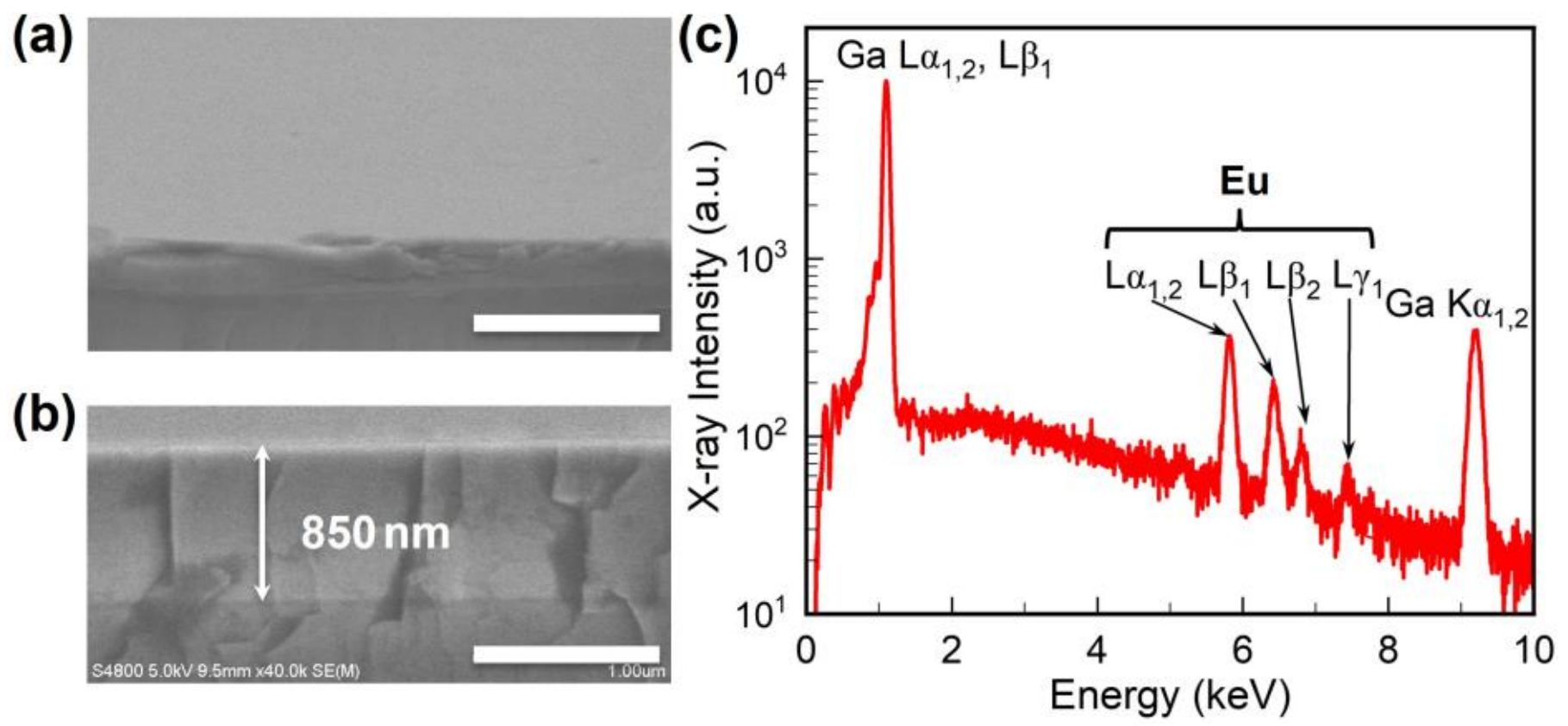

**Figure 1.** Structural and compositional characterization of a GaN:Eu epilayer with an Eu concentration of $3.7\times10^{19}$ $cm^{-3}$. (a), (b) 70°-tilted view and cross-sectional SEM micrographs. Scale bars: 1 μm. (c) EDS spectrum.

High-quality GaN:Eu epilayers were grown by PA-MBE on 2-inch *c*-plane GaN-on-sapphire templates. Figure 1 summarizes the structural and compositional characterization of a representative sample doped at $3.7\times10^{19}$ $cm^{-3}$, selected for all the subsequent optical measurements. The 70°-tilted and cross-sectional SEM micrographs in Figs. 1(a) and (b) show that the 850-nm-thick GaN:Eu epilayer exhibits a smooth surface sporadically decorated with V-shaped pits. The energy-dispersive X-ray spectroscopy spectrum in Fig. 1(c) exhibits distinct Eu *L*-series characteristic peaks, confirming effective incorporation into the GaN host lattice. Secondary ion mass spectrometry (SIMS, Suppl. Fig. S1) shows that Eu concentration increases linearly with effusion cell beam flux, saturating at approximately $2.7\times10^{20}$ $cm^{-3}$, with higher flux triggering Eu-enriched surface precipitates. X-ray diffraction (XRD) $2\theta$-$\omega$ scans and ω rocking curves of the (002) and (102) planes are nearly identical for GaN:Eu and undoped GaN epilayers (Suppl. Fig. S2), with rocking curve FWHMs of 100—250 arcsec, indicating excellent crystalline quality.

Room-temperature PL spectra under 325 nm above-bandgap excitation are presented in Fig. 2(a) for the samples with varying Eu doping levels. Prominent band-edge emission from undoped GaN at 3.412 eV is strongly suppressed in GaN:Eu, indicating effective host-to-$Eu^{3+}$ energy transfer. At a low Eu concentration of $3.5\times10^{18}$ $cm^{-3}$, broad yellow luminescence from vacancy-impurity pairs[36-38] is observed, alongside four peaks at 3.319, 3.229, 3.138, and 3.045 eV with a uniform energy spacing of 92±2 meV, matching the 1st through 4th-order LO phonon replicas of the band-edge exciton. This observation indicates that $Eu^{3+}$ doping enhances exciton-LO phonon coupling via the Fröhlich interaction[39, 40]. As Eu concentration exceeds $1.0\times10^{19}$ $cm^{-3}$, the blue and yellow luminescence of GaN host is fully suppressed, and only sharp $Eu^{3+}$ emission lines are observed, reflecting efficient energy transfer[28, 41]. The high-resolution PL spectrum in Fig. 2(b) shows that these lines arise primarily from $^5D_0\rightarrow{}^7F_J$ ($J$=0–3) and $^5D_1\rightarrow{}^7F_J$ ($J$=1–4) transitions. Notably, the $^5D_0\rightarrow{}^7F_2$ and $^5D_1\rightarrow{}^7F_4$ red emission lines account for 80% of integrated intensity, demonstrating the potential for high-purity red light applications. The inset shows visible red-light emission from the $3.7\times10^{19}$ $cm^{-3}$ sample. Further increases in Eu concentration diminish $Eu^{3+}$ luminescence owing to agglomeration and concentration quenching[42].

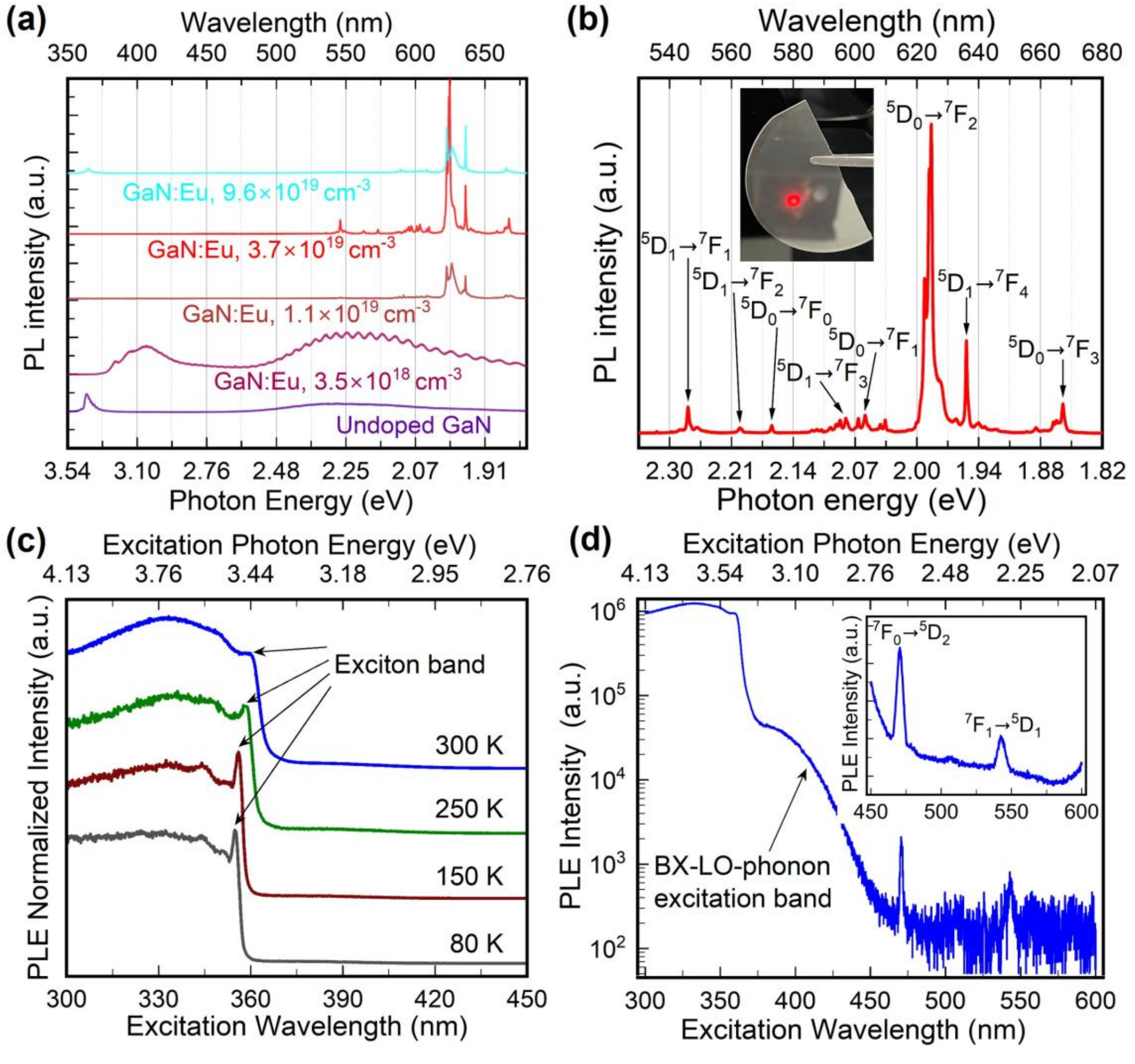

**Figure 2.** Photoluminescence (PL) and PL excitation (PLE) characterizations of GaN:Eu epilayers. (a) PL spectra for samples with varying Eu doping concentrations; the $3.7\times10^{19}$ $cm^{-3}$ sample (red

curve) is used for (b)—(d). (b) High-resolution PL spectrum for $Eu^{3+}$ transition analysis; inset: PL emission photograph. (c) Temperature-dependent PLE spectra monitored at 622 nm (linear scale). All intensities are normalized to the signal at 325 nm. (d) Room-temperature semi-log PLE spectrum resolving the BX-LO-phonon polaron excitation band (370–450 nm) and resonant excitation peaks at 470.5 nm ($^7F_0 \rightarrow {}^5D_2$) and 543.0 nm ($^7F_1 \rightarrow {}^5D_1$).

To clarify the role of excitons in energy transfer, we performed temperature-dependent PL excitation (PLE) spectroscopy monitored at the 622 nm $^5D_0 \rightarrow {}^7F_2$ emission [Fig. 2(c)]. A sharp excitonic excitation peak is clearly resolved at low temperatures. Varshni fitting yields $E_X$= 3.499 eV, $\alpha$= 0.691 meV/K, and $\beta$= 850.5 K[43]. Gradual quenching of the peak with rising temperature demonstrates that exciton dissociation suppresses $Eu^{3+}$ excitation, confirming direct exciton-mediated energy transfer, consistent with prior PA-MBE results[33, 34]. In contrast, previous work on OMVPE-grown GaN:Eu reported the absence of low-temperature excitonic features and proposed that excitons do not participate in host-to-$Eu^{3+}$ energy transfer[32]. Such contradictory energy-transfer mechanisms originate from structurally distinct luminescent centers in PA-MBE and OMVPE samples, as validated by CEES characterizations[44-46].

The room-temperature PLE spectrum plotted on a semi-log scale in Fig. 2(d) resolves a broad sub-bandgap excitation band spanning 370–450 nm and two sharp resonant peaks at 470.5 and 543.0 nm. This broad sub-bandgap band, matching the blue luminescence band in Fig. 2(a), originates from overlapping bound-exciton polaron transitions with $n$=1–7 LO phonons, merged by room-temperature homogeneous and inhomogeneous broadening. Fitting reveals intensities decreasing with phonon order, ~5% of the exciton peak at $n$=1, 1% at $n$=5, and exponentially decay to baseline for $n \geqslant 7$, consistent with the inherently low oscillator strength of second-order phonon-assisted transitions[47]. The two sharp peaks arise from resonant $^7F_0 \rightarrow {}^5D_2$ and $^7F_1 \rightarrow {}^5D_1$ transitions[48, 49], with 0.1—0.2% relative intensity due to small optical cross section of parity-forbidden 4$f$ transitions. Direct $^5D_2$ excitation populates the $^5D_0$ state via the $^5D_1$ intermediate more efficiently than direct $^5D_1$ excitation, in agreement with prior reports[50]. No direct $^5D_0$ resonant excitation or phonon-coupled resonant excitation is observed here[45].

To resolve the fine structure of exciton-phonon coupling and identify intermediate energy transfer states, we performed temperature-dependent PL measurements from 5 to 300 K, as summarized in Fig. 3. At cryogenic temperatures, GaN:Eu yields both $Eu^{3+}$ intra-4*f* transition emission and GaN host band-edge exciton emission with intense LO-phonon replicas. Figure 3(c) compares the band-edge region of GaN:Eu and undoped GaN. Undoped GaN is dominated by type-A free exciton ($FX_A$) at 3.491 eV and its shallow donor-bound counterpart ($BX_1$, 3.486 eV), with weaker type-B free exciton ($FX_B$, 3.500 eV)[51-53], bound exciton ($BX_2$, 3.482 eV) and two-electron satellite (TES, 3.471 eV) peaks. In GaN epilayers, $BX_1$ and $BX_2$ are generally attributed to excitons bound to unintentionally incorporated impurities (*e.g.,* Si, O) and nitrogen vacancies ($V_N$)[54, 55]. In contrast, GaN:Eu band-edge emission is dominated by a prominent bound exciton peak $BX_3$ at 3.478 eV, alongside residual $BX_1$ and a weak $FX_A$ shoulder. $Eu^{3+}$ doping enables efficient trapping of free excitons, leading to pronounced quenching of free exciton emission. Deconvolution reveals an additional bound exciton $BX_4$ at 3.473 eV and a TES peak at 3.463 eV associated with either $BX_3$ or $BX_4$. These characteristic exciton states are likely localized modes at $Eu^{3+}$-related lattice sites[56, 57], where free excitons are trapped to form bound excitons and create an efficient energy transfer channel. Specifically, one of the six 4*f* electrons of $Eu^{3+}$ participates in the two-electron ensemble of the bound exciton, facilitating nonradiative energy transfer to the $^5D_J$ state of $Eu^{3+}$.

The most striking result is the observation of up to seven orders of LO-phonon replicas of $BX_3$, shown in Fig. 3(d). All the replicas have uniform energy spacing of 92.0±0.7 meV, matching the characteristic GaN LO-phonon energy. As shown in Fig. 3(e), the PLE spectrum monitored at 622 nm $Eu^{3+}$ emission shows nearly perfect overlap with the host PL spectrum at 5K, spanning the entire $BX_3$ zero-phonon line and all its LO-phonon replicas. This one-to-one correspondence provides direct, unambiguous evidence that $BX_3$ excitation proceeds exclusively via these bound-exciton polaron states, rather than through isolated deep defect levels. The $5^{th}$-order LO-phonon replica energy of 3.019 eV is identical to the $Eu^{3+}$ $^7F_0 \rightarrow {}^5D_3$ transition energy [58]. This precise resonance represents the defining signature of polaron-mediated energy transfer. The $BX_3$ exciton cascades downward via sequential LO-phonon emission and resonantly transfers energy to $Eu^{3+}$ upon reaching the 5×LO-polaron state, exciting the ion from $^7F_0$ ground state to $^5D_3$ excited state.

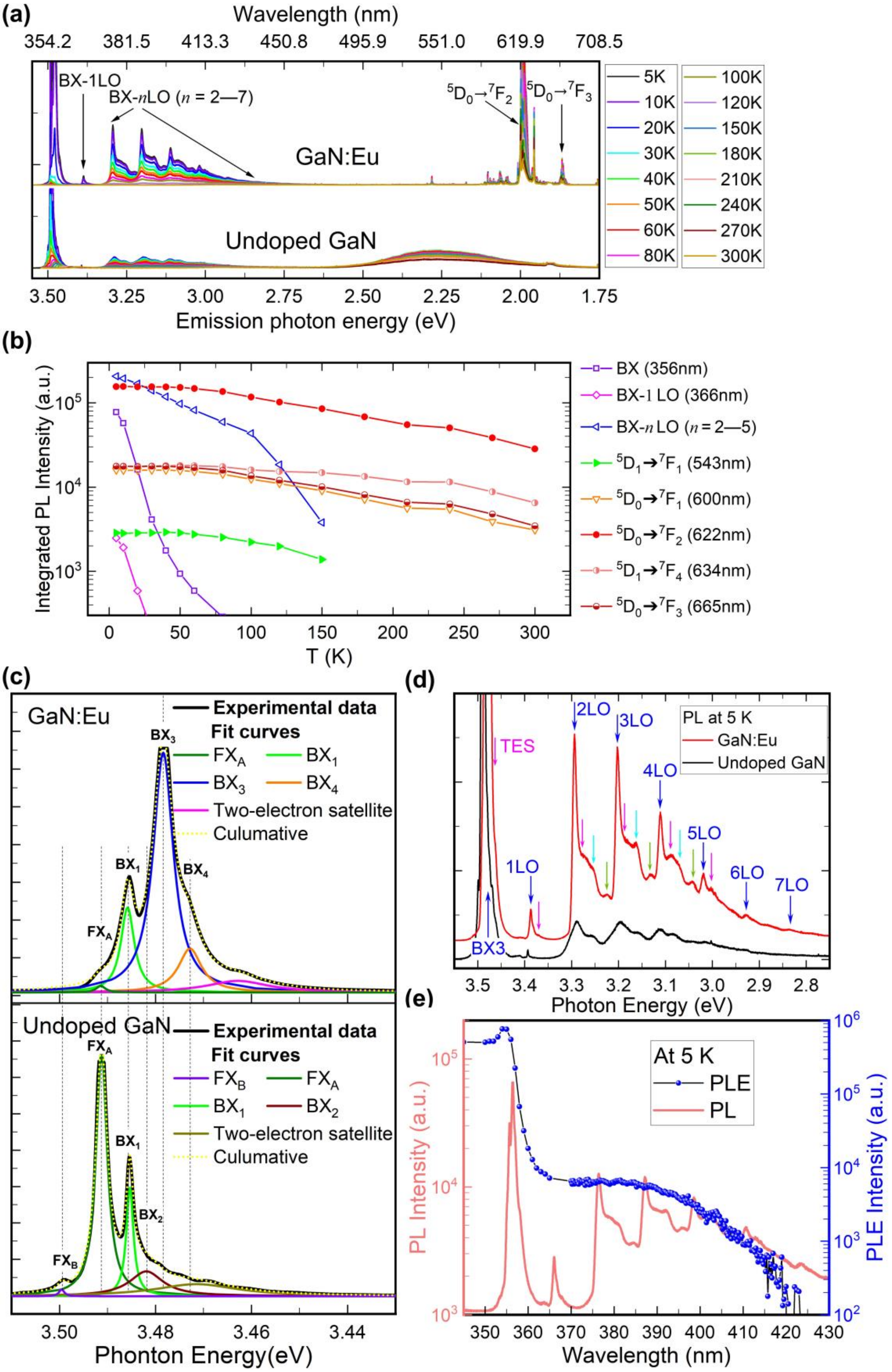

**Figure 3.** Temperature-dependent PL. (a) Comparison between GaN:Eu and undoped GaN. (b) Integrated PL intensity of GaN:Eu versus temperature. (c) Comparison of the exciton peaks of GaN:Eu and undoped GaN at 5 K. (d) LO-phonon replicas of the $BX_3$ bound exciton in GaN:Eu. (e) Comparison of PLE and PL spectra of GaN:Eu at 5 K.

Further evidence for resonant energy transfer comes from the temperature evolution of LO-phonon replica intensities. Classical lattice dynamics theory predicts that the multi-phonon radiative line intensities obey a Poisson distribution, with the relative intensity profile governed by the *Huang-Rhys factor S* and invariant with temperature, provided that $k_B T \ll \hbar\omega_0$ is satisfied and phonons are emitted independently under thermodynamic equilibrium[59-61]. The *S* factor extracted from the intensity ratio of $1^{st}$-order replica to zero-phonon line is 0.023 for GaN:Eu and 0.016 for undoped GaN, consistent with the previous reports[59, 62]. However, the measured temperature dependence deviates dramatically from equilibrium predictions. As shown in Fig. 3(b), the relative intensities of $2^{nd}$—$5^{th}$-order LO-phonon replicas grow markedly with respect to the $BX_3$ zero-phonon line and $1^{st}$-order replica as temperature rises, while the higher-order replicas exhibit pronounced quenching. This anomalous behavior falls outside equilibrium phonon sideband theory, and instead arises from non-equilibrium phonon emission driven by directional resonant energy transfer to $Eu^{3+}$. Elevated temperatures accelerate exciton-phonon scattering, promoting cascaded LO-phonon emission and populating the intermediate *n*=2–5 polaron states. These intermediate states are efficiently depopulated via resonant energy transfer to the $Eu^{3+}$ $^5D_3$ state, which in turn suppresses radiative emission from higher-order replicas. This non-equilibrium polaron dynamics constitutes a definitive signature of active energy transfer, as opposed to passive phonon sideband emission.

To investigate $Eu^{3+}$ excited-state dynamics and identify luminescent sites, we performed time-resolved PL measurements at 80 and 300 K for three characteristic emission lines, $^5D_0 \rightarrow {}^7F_2$ (622 nm), $^5D_0 \rightarrow {}^7F_3$ (665 nm) and $^5D_1 \rightarrow {}^7F_4$ (634 nm), with decay transients shown in Fig. 4. All the transients are well fitted by the biexponential decay model in Equation (1), yielding a temperature-insensitive slow component and a temperature-sensitive fast component (full fitting parameters in Table 1).

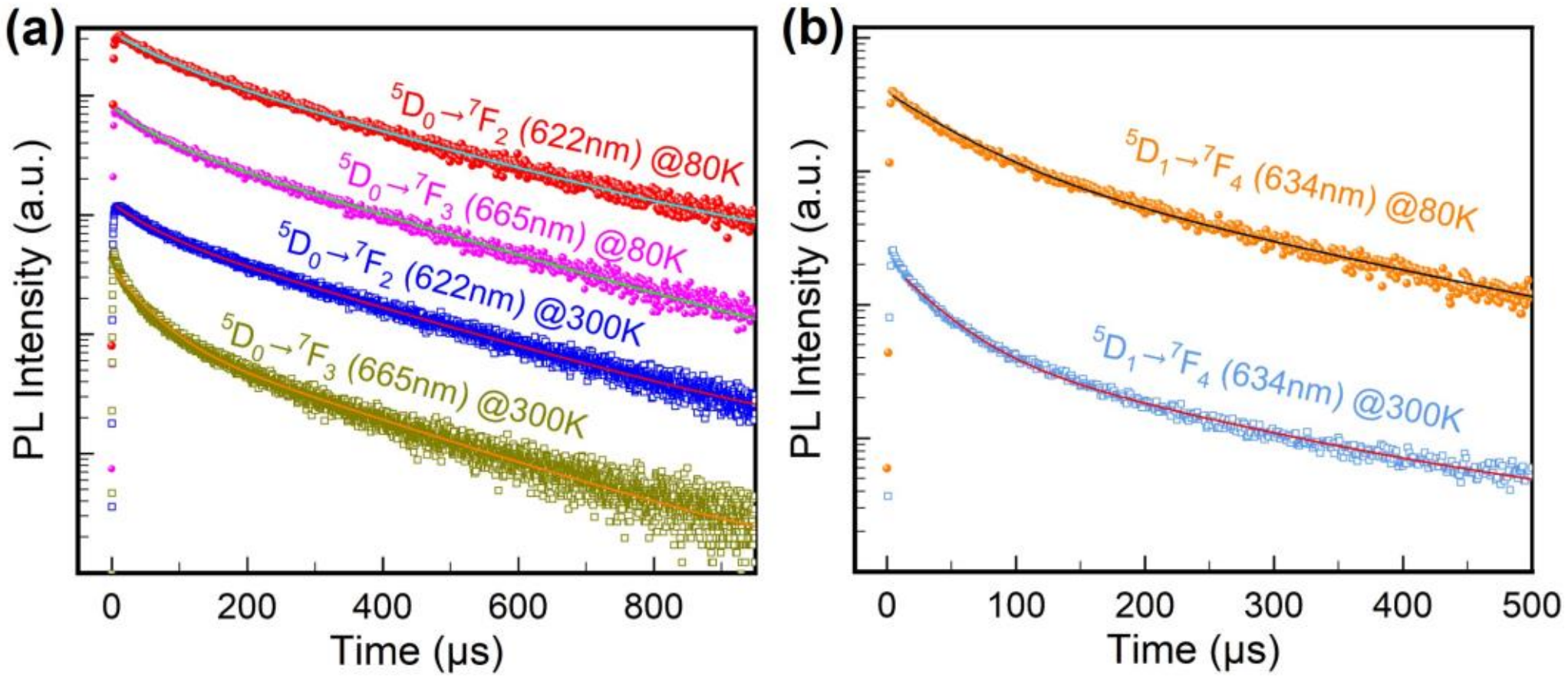

**Figure 4.** Time-resolved PL (TRPL) of the GaN:Eu epilayer. (a) TRPL decay transients for $^5D_0 \rightarrow ^7F_2$ (622 nm) and $^5D_0 \rightarrow ^7F_3$ (665 nm) at 80 and 300 K. (b) TRPL decay transients for $^5D_1 \rightarrow ^7F_4$ (634 nm) at 80 and 300 K. The spectral window is set to 2 nm.

For the $^5D_0$ manifold, the slow decay times of $^5D_0 \rightarrow ^7F_2$ and $^5D_0 \rightarrow ^7F_3$ are indistinguishable within experimental error, averaging 252 ± 5 μs. This temperature-invariant component is assigned to radiative decay of $^5D_0$ to $^7F_J$ ($J$=0—3) manifolds. Per Equation (2), the $^5D_0 \rightarrow ^7F_2$ transition dominates the overall decay dynamics, as its steady-state integrated intensity exceeds that of all other $^5D_0$ emissions by over one order of magnitude. The much longer intrinsic radiative lifetime of the $^5D_0 \rightarrow ^7F_3$ transition (>1 ms) renders it a more sensitive probe of weak nonradiative pathways and subtle $Eu^{3+}$ site variations. Equation (3) accounts for the total $^5D_0$ decay contribution from all the distinct $Eu^{3+}$ lattice sites. Previous CEES studies have reported radiative lifetimes ranging from 207 to 290 μs for various $Eu^{3+}$ sites [25, 28, 29]. The nearly identical radiative lifetimes for both $^5D_0$ transitions demonstrate that a single $Eu^{3+}$ site dominates the luminescence.

**Table 1.** Biexponential fit results for the fast and slow decay times. The detected initial rise time is approximately 0.90 μs for all measurements.

| Emission line | Temperature | $\tau_1$ (μs) | $A_1$ | $\tau_2$ (μs) | $A_2$ | $\frac{A_1}{A_1 + A_2}$ |
|---|---|---|---|---|---|---|
| $^5D_0 \rightarrow ^7F_2$ (622 nm) | 80 K | 67.35 | 599.75 | 255.32 | 1286.1 | 31.80% |
| | 300 K | 55.77 | 554.62 | 251.39 | 1179.9 | 31.98% |
| $^5D_0 \rightarrow ^7F_3$ (665 nm) | 80 K | 67.08 | 1674.3 | 254.20 | 1433.0 | 53.88% |
| | 300 K | 43.13 | 831.26 | 249.45 | 408.39 | 67.06% |
| $^5D_1 \rightarrow ^7F_4$ (634 nm) | 80 K | 38.22 | 368.69 | 169.41 | 652.76 | 36.09% |
| | 300 K | 22.54 | 3207.0 | 167.36 | 1247.5 | 71.99% |

$$I(t) = A_1 exp\left(-\frac{t}{\tau_1}\right) + A_2 exp\left(-\frac{t}{\tau_2}\right) \quad (1)$$

$$I_i \propto \frac{N^*_{Eu,i}}{\tau_{decay,i}} \quad (2)$$

$$\sum_i \frac{N_i^{^5D_0}}{\tau_i^{^5D_0}} = \sum_i \frac{N_i^{^5D_0}}{\tau_i^{^5D_0\rightarrow{}^7F_2}} + \sum_i \frac{N_i^{^5D_0}}{\tau_i^{^5D_0\rightarrow{}^7F_3}} + \sum_i \frac{N_i^{^5D_0}}{\tau_i^{nr'}} \quad (3)$$

$$\sum_i \frac{N_i^{^5D_1}}{\tau_i^{^5D_1}} = \sum_i \frac{N_i^{^5D_1}}{\tau_i^{^5D_1\rightarrow{}^5D_0}} + \sum_i \frac{N_i^{^5D_1}}{\tau_i^{^5D_1\rightarrow{}^7F_1}} + \sum_i \frac{N_i^{^5D_1}}{\tau_i^{^5D_1\rightarrow{}^7F_4}} + \sum_i \frac{N_i^{^5D_1}}{\tau_i^{nr''}} \quad (4)$$

Minor perturbations to the local crystal field, by contrast, strongly modify nonradiative processes such as $Eu^{3+}$ cross-relaxation and coupling of the $^5D_0$ state to impurities or trap states. At 80 K, the fast decay times are nearly equivalent (~67 μs) for both $^5D_0$ transitions, but diverge substantially at 300 K, 55.77 μs for $^5D_0\rightarrow{}^7F_2$ and 43.13 μs for $^5D_0\rightarrow{}^7F_3$. The moderate reduction for $^5D_0\rightarrow{}^7F_2$ reflects thermally activated nonradiative depopulation of the dominant luminescent site, while the steeper decline for $^5D_0\rightarrow{}^7F_3$ signals activation of an additional nonradiative channel associated with a minority site. This interpretation is corroborated by the normalized fast-component amplitude ratio $A_1/(A_1+A_2)$, a quantitative metric of nonradiative contribution. The ratio is approximately 32% for the dominant site probed via $^5D_0\rightarrow{}^7F_2$, and exceeds 50% when the minority site is sampled through $^5D_0\rightarrow{}^7F_3$.

For the $^5D_1$ manifold, $^5D_1\rightarrow{}^7F_4$ transients yield a nearly temperature-independent slow decay time of $168\pm2$ μs, while the fast component shortens from 38.22 μs at 80 K to 22.54 μs at 300 K. This acceleration indicates thermally enhanced $^5D_1\rightarrow{}^5D_0$ internal conversion and other nonradiative depopulation pathways. Equation (4) describes the total $^5D_1$ decay from all the $Eu^{3+}$ sites. The primary $^5D_1$ depopulation channel is the $^5D_1\rightarrow{}^5D_0$ nonradiative transition with a reported characteristic time constant of ~2 μs[34, 50], which is below the temporal resolution of our setup given the ~1 μs excitation pulse width. The observed slow decay time is governed by the radiative $^5D_1\rightarrow{}^7F_4$ decay, as its steady-state integrated intensity far exceeds that of $^5D_1\rightarrow{}^7F_1$. The weaker overall PL intensity of $^5D_1\rightarrow{}^7F_4$ relative to $^5D_0\rightarrow{}^7F_2$ stems from the lower steady-state population of the $^5D_1$ excited state.

High-resolution cryogenic PL reveals detailed crystal-field splitting of $Eu^{3+}$ emission lines, providing insight into local site structure. Figure 5 shows that the $^5D_0 \rightarrow {}^7F_2$ transition splits into five distinct peaks, and the $^5D_0 \rightarrow {}^7F_3$ transition splits into seven peaks at cryogenic temperatures. This observation confirms that the point-group symmetry of the $Eu^{3+}$ site is lower than the $C_{3v}$ symmetry of GaN, implying that the photoactive site deviates from ideal Ga substitutional sites or couple with adjacent point defects. The $^7F_J$ state of $Eu^{3+}$, with a $(2J+1)$-fold degeneracy from the total angular momentum of 4*f* electrons, can have its degeneracy partially or totally lifted by crystal-field effect, with a lower coordination symmetry promoting mixing between 4*f* and 5*d* orbitals to enhance intra-4*f* transitions[63, 64]. For $RE^{3+}$ ions substituting $Ga^{3+}$ sites, the 5-fold degenerated $^7F_2$ state splits into one single and two doubly degenerate crystal-field levels, yielding three PL peaks with energy separations <100 $cm^{-1}$ [26, 65]. Here we observe five peaks for $^5D_0 \rightarrow {}^7F_2$ located at 620.02, 620.72, 621.62, 622.48, and 623.88 nm, with energy separations of 18.19, 23.32, 22.22, and 36.05 $cm^{-1}$ ($m_J = 0, \pm1, \pm2$). These results unambiguously demonstrate that the dominant $Eu^{3+}$ luminescence center deviates from relaxed $Ga^{3+}$ substitutional sites, likely due to the larger ionic radius and smaller Pauling electronegativity of $Eu^{3+}$ compared with $Ga^{3+}$, which distorts the surrounding nitrogen tetrahedron.

To deduce the $Eu^{3+}$ site configuration, we consider established defect models. Previous work on Eu/O co-doped GaN proposed $RE^{3+}$ ions at Ga sites coordinated with two adjacent O atoms and two N atoms[18, 19, 28, 66], a configuration that breaks $C_{3v}$ symmetry. However, our SIMS analysis shows oxygen impurity concentrations below the detection limit, ruling out this O-mediated configuration. $RE_{Ga}$-$V_N$ pairs have the largest calculated binding energy (1.0 eV for Eu), and are thermally stable up to 1000 °C[67]. We therefore turn to the two $RE_{Ga}$-$V_N$ pair configurations proposed by B. Mitchell *et al*[20, 26, 68, 69]. $C_{3v}$ symmetry is preserved when a $V_N$ lies along the *c*-axis above $Eu^{3+}$, while symmetry is reduced to $C_{1h}$ when a $V_N$ occupies one of the three equivalent off-center positions. Given the observed reduced symmetry, the off-axis configuration is the most plausible luminescent site [Fig. 5(c)]. The consistent splitting behavior under varying excitation wavelengths and power levels [Suppl. SFig. 5] further validates the single dominant luminescence site identified by lifetime analysis.

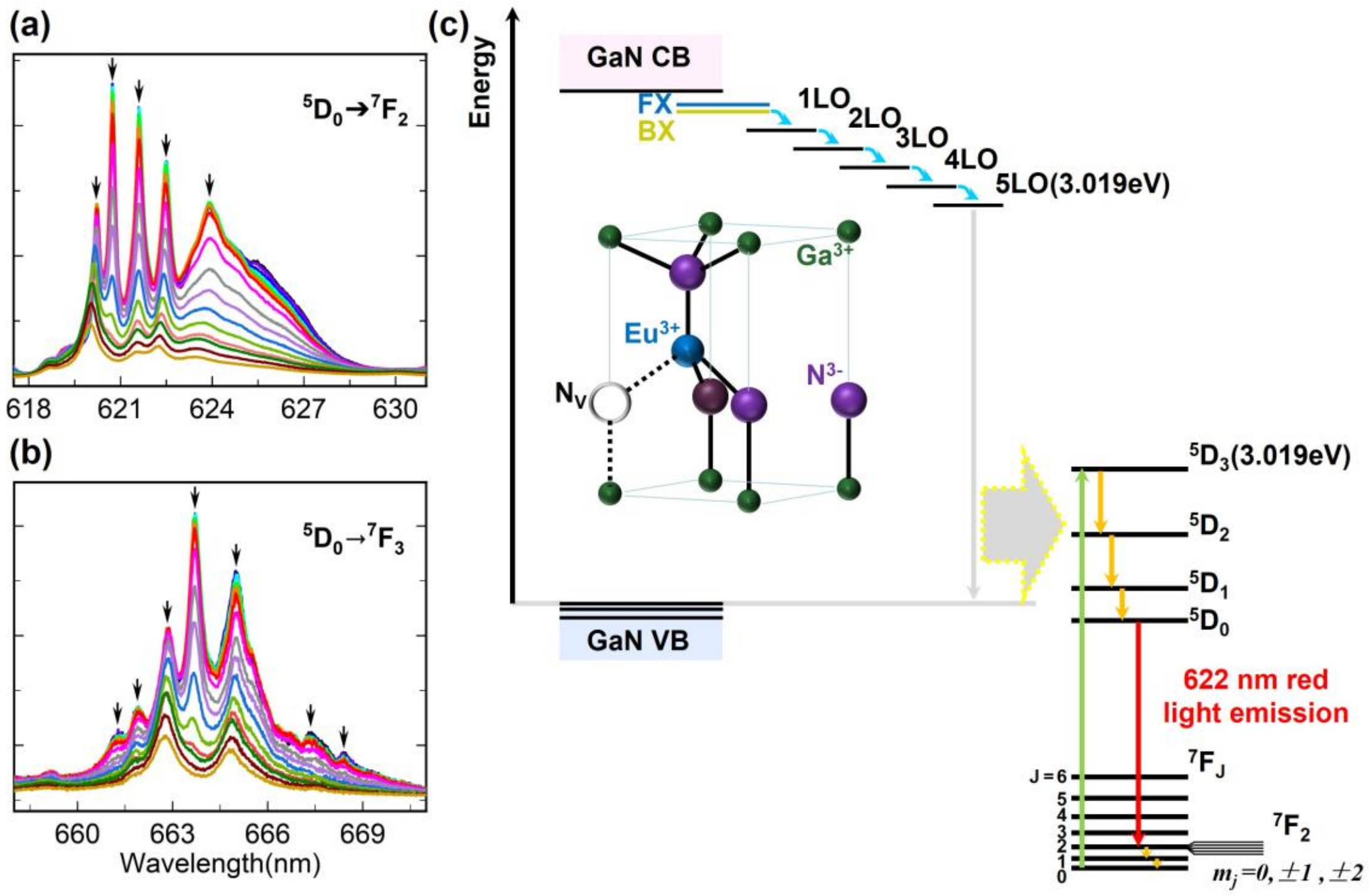

**Figure 5.** Crystal-field splitting of $Eu^{3+}$ luminescence and the energy transfer mechanism in GaN:Eu. (a) 5-fold splitting of the $^5D_0\to{}^7F_2$ transition. (b) 7-fold splitting of the $^5D_0\to{}^7F_3$ transition. (c) Schematic of the energy transfer pathway from the GaN host to $Eu^{3+}$.

Putting these observations together, we propose a complete cascading LO-phonon-mediated resonant energy transfer mechanism model, illustrated schematically in Fig. 5(c). Above-bandgap excitation generates free excitons in the GaN host, which are rapidly trapped at $Eu^{3+}$-related sites to form $BX_3$ bound excitons. These excitons interact strongly with LO phonons via the Fröhlich interaction, cascading downward through sequential polaron states by emitting LO phonons. When the exciton-phonon polaron reaches the $5^{th}$-LO-phonon state at 3.019 eV, it resonantly transfers its energy to $Eu^{3+}$, exciting the ion from the $^7F_0$ ground state to the $^5D_3$ excited state. The excited $Eu^{3+}$ then undergoes rapid nonradiative multi-phonon relaxation through the $^5D_J$ manifold, ultimately populating the metastable and lowest-lying excited state $^5D_0$. Finally, radiative $^5D_0\to{}^7F_2$ transition produces the characteristic 622 nm high-purity red emission. This mechanism clarifies the long-standing debate about host-to- $Eu^{3+}$ energy transfer, and the observed non-equilibrium phonon dynamics provide direct evidence for directional energy flow. The identification of the dominant off-axis $Eu_{Ga}$-$V_N$ luminescent site also offers a clear target for defect engineering to further enhance emission efficiency.

In summary, we have grown high-crystalline-quality GaN:Eu epilayers free of Eu-rich segregated precipitates by systematic tuning of epitaxial conditions. Above-bandgap optical excitation produces bright room-temperature red emission, with strong suppression of GaN host band-edge and defect luminescence. We elucidate the GaN-host-to-$Eu^{3+}$ energy transfer mechanism for high-color-purity 622 nm red emission. Cascading polarons formed by bound excitons and $n$×LO phonons ($n$=1—7) enable resonant excitation of the $Eu^{3+}$ $^7F_0 \rightarrow {^5D_3}$ transition. Both the 5-fold splitting of $^5D_0 \rightarrow {^7F_2}$ and 7-fold splitting of $^5D_0 \rightarrow {^7F_3}$ validate reduced $Eu^{3+}$ site symmetry, consistent with off-axis $Eu_{Ga}$-$V_N$ complex formation. This work establishes a comprehensive physical framework for the design and realization of high-performance red-light emitters based on Eu-doped GaN.

**Additional information**

Supplementary information is available online.

**ACKOWLEDGMENTS**

The authors acknowledge Prof. Shi-jie Xu (xusj@fudan.edu.cn) at Fudan University for discussion and suggestions on the excitons and polarons in GaN, and Prof. Jianxun Liu (jxliu2018@sinano.ac.cn) at SINANO for manuscript revision. The authors are also grateful for the technical support from Nano Fabrication Facility, Platform for Characterization and Test and Nano-X from SINANO, CAS, and particularly acknowledge Dr. Xiaotian Ge (xtge2017@sinano.ac.cn) for the assistance in PL setup. This study is supported by the CAS Young Talent program under Grant No. E2290106, and National Natural Science Foundation of China (Grant No. 62274173).

**AUTHOR DECLARATIONS**

**Conflict of Interest**

The authors have no conflicts of interest to disclose.

## Author Contributions

S. Fan conceived the research and designed the experiment. S. Fan and Q. Sun supervised the overall research activity. S. Fan, Y. Xu, and Y. Yin conducted the MBE growth. S. Fan, Y. Xu, and S. Dong performed the PL experiment and analysis. S. Fan, Y. Xu, Z. Su and S. Chen conducted the PL transient decay experiment. S. Fan, Y. Xu, D. Tu, and X. Chen conducted the power-dependent PL experiment. S. Fan wrote the manuscript with comments and revisions from all other authors.

## DATA AVAILABILITY

The data that supports the findings of this study is available within the article and its supplementary information. Additional data is available from the corresponding author upon reasonable request.

# METHODS

## Materials

GaN:Eu thin films were grown on commercial 4-μm-thick GaN-on-Sapphire templates (SinoVio Semiconductor Technol. *Co. Ltd.*) using a custom-built PA-MBE system (Fermion Instruments *Co., Ltd.*). The templates were fabricated via organometallic vapor phase epitaxy (OMPVE) on sapphire substrates, incorporating a 25-nm sputtered and post-annealed AlN nucleation layer. The ultra-high vacuum (UHV) growth chamber achieved a base pressure below $1\times10^{-10}$ Torr. The epitaxial growth was performed at a substrate temperature of 750 °C with a nominal growth rate of 200 $nm\cdot h^{-1}$. High-purity gallium (7N5) was evaporated from a standard SUMO effusion cell heated to 1050–1150 °C, maintaining a beam-flux equivalent pressure (BEP) of $1 \times 10^{-}$

$^{7}$ to 5 × $10^{-7}$ Torr. Europium doping was realized using high-purity Eu chunks (4N) loaded in a conical pyrolytic boron nitride crucible, with thermal evaporation of Eu conducted at 300–430 °C. A Bayard–Alpert ionization gauge was integrated into the chamber to monitor the BEP of both Ga and Eu atomic beams. Active nitrogen species were generated by a 13.56 MHz inductively coupled plasma source. High-purity $N_2$ gas (6N) was supplied at a flow rate of 0.2–1.0 standard cubic centimeters per minute, precisely regulated by a mass flow controller. The plasma power was adjusted between 300 W and 450 W to optimize nitrogen activation efficiency. All GaN:Eu epilayers were grown atop a 100-nm-thick undoped GaN buffer and capped with a 5-nm GaN layer to suppress surface oxidation of Eu. Undoped control samples with an identical layer structure and thickness were grown under the same conditions for comparison. In addition, a GaN:Eu/GaN:Si multilayer structure consisting of alternating Eu-doped and Si-doped GaN layers was separately for quantitative calibration of Eu doping concentration (Suppl. Fig. S1).

**Characterizations and measurements**

Energy-dispersive X-ray spectroscopy (EDS) was performed on a field-emission scanning electron microscope (FE-SEM, Nova NanoSEM 450) operated at an accelerating voltage of 15 keV and a beam current of 1.0 nA. Time-of-flight secondary ion mass spectrometry (TOF-SIMS) depth profiling was carried out using a TOF.SIMS 5-100 system (ION-TOF GmbH, Münster, Germany). For positive ion detection, a 30 keV $O_2^-$ beam served as the primary analysis beam, while a 1 keV $Cs^+$ beam was used for sputter etching to expose underlying layers. High-resolution X-ray diffraction (HR-XRD) measurements, including 2θ-ω scans and ω-rocking curve scans, were collected on a PANalytical X'pert3 MRD diffractometer to evaluate the crystal quality and structural properties of the epilayers. The diffractometer was equipped with a Cu-$K_\alpha$ radiation source (λ = 1.54056 Å) operated at 40 kV and 45 mA, a Ge (220) symmetric four-bounce monochromator, and a scintillator detector. Raman scattering measurements were acquired on a wafer-mapping Raman system (Model R1, Suzhou Lumixplore *Co., Ltd.*) equipped with a 320 mm-focal-length spectrometer and an 1800

grooves·mm$^{-1}$ grating. A 532 nm laser (30 mW output power) was focused to a spot diameter of ~0.8 μm through a 100× objective lens (NA = 0.8).

Temperature-dependent photoluminescence (PL) measurements were performed using a 325 nm continuous-wave Cd-He laser as the excitation source, with an iHR550 spectrometer for spectral collection. The experimental parameters were strictly controlled with a spectral step size of 0.023 nm, an excitation laser power density of 67 mW·cm$^{-2}$, a laser spot diameter of <3.65 mm, and an integration time of 1 s per data point. For accurate temperature regulation, samples were mounted on a copper heat sink using a thin, uniform layer of high-thermal-conductivity paste to minimize interfacial thermal resistance. The sample assembly holder was then loaded into a high-vacuum helium-flow cryostat, enabling measurements over a wide temperature range.

To confirm that the five characteristic emission peaks assigned to the $^5D_0 \rightarrow {}^7F_2$ transition arise from a single type of $Eu^{3+}$ photoactive site, power-dependent PL measurements at 10 K were conducted using a pulsed 355 nm laser (MPL-H-355, Changchun New Industries Optoelectronics Tech. *Co., Ltd.*). The laser was operated at a power density of 470 mW·cm$^{-2}$, a pulse width of 1.3 ns, and a repetition rate of 4—7 kHz. The excitation power was varied from 2.6% to 100% of the maximum output using a set of neutral-density filters, and the corresponding spectral evolution is shown in Suppl. Fig. S5.

Photoluminescence excitation (PLE) measurements were carried out using a xenon lamp (EQ-99X-QZ-S, 0.75 W full spectral output) coupled to an iHR320 spectrometer. Monochromatic excitation light was scanned at a step size of 0.1 nm across the target wavelength range. Time-resolved PL (TRPL) measurements were performed at room temperature and 80 K, using a 60 W microsecond xenon flash lamp (μF2, HAMAMATSU Photonics K. K.) as the pulsed excitation source (full spectral output power ~100 μW). Excitation light was monochromatized by a dual-grating excitation monochromator (DTGM325-X) and fixed at 325 nm. The flash lamp exhibited a pulse width of ~1 μs and a repetition frequency of 100 Hz, with both excitation and emission spectral bandwidths set to 1 nm. Transient luminescence signals were dispersed by a dual-grating emission monochromator (DTGM325-M) and detected by a cooled red-sensitive photomultiplier tube (PMT-900). All decay transients were acquired in multi-channel scaling mode, and TRPL decay curves were individually recorded at the three characteristic $Eu^{3+}$ emission peaks of 622 nm, 634 nm, and 665 nm.